\documentclass{article}
\usepackage{spconf,amsmath,graphicx}
\usepackage{color}
\usepackage{multirow}
\usepackage{balance}
\usepackage{fancyhdr}
\newcommand{\newvspace}{\vspace{-0.40cm}}
\newcommand{\imagevspace}{\vspace{-0.45cm}}

\title{An ensemble framework of voice-based emotion recognition system for films and TV programs}
\makeatletter
\def\name#1{\gdef\@name{#1\\}}
\makeatother
\name{{\em Fei Tao$^{1\ast}$ \thanks{$\ast$This work was done during the author's summer internship at Alibaba Group (U.S.) Inc.}, Gang Liu$^2$, Qingen Zhao$^2$\vspace{-0.4cm}}}
\address{1. Multimodal Signal Processing (MSP) Lab, The University of Texas at Dallas, Richardson TX\\
	2. Institute of Data Science and Technology (iDST)-Speech, Alibaba Group (U.S.) Inc.\\
	{\footnotesize \tt \hspace{-0.1cm}fxt120230@utdallas.edu,g.liu@alibaba-inc.com,qingen.zqe@alibaba-inc.com} \vspace{-0.5cm}}

\begin{document}
%
%
\maketitle
\begin{abstract}
Employing voice-based emotion recognition function in \emph{artificial intelligence} (AI) product will improve the user experience. Most of researches that have been done only focus on the speech collected under controlled conditions. The scenarios evaluated in these research were well controlled. The conventional approach may fail when background noise or non-speech filler exist. In this paper, we propose an ensemble framework combining several aspects of features from audio. The framework incorporates gender and speaker information relying on multi-task learning. Therefore it is able to dig and capture emotional information as much as possible. This framework is evaluated on \emph{multimodal emotion challenge} (MEC) 2017 corpus which is close to real world. The proposed framework outperformed the best baseline system by 29.5\% (relative improvement).
\end{abstract}
\begin{keywords}
multi-task learning, attention model, ensemble framework, deep learning, emotion recognition
\end{keywords}
\imagevspace
\section{Introduction}
\label{sec:intro}
\newvspace
Humans are emotional creatures. People desire reaction from others according to their emotion \cite{Busso_2004}. \emph{Artificial intelligence} (AI) system will be more like human beings when it is able to do such reaction. This capability relies on recognizing emotion. Emotion recognition is therefore very important in AI product, since it will make \emph{human-computer interface} (HCI) more friendly and improve the user experience. 

Emotion recognition system based on audio (which can also be seen as voice-based) has very low requirement for hardware, even though multimodal speech processing can improve speech related system performance \cite{tao_2014,tao_2017}. Therefore the audio based emotion recognition system is easier to be employed on AI product \cite{Busso_2004,Schuller_2003} than other means. However, current voice-based AI products, such as Siri, Google voice search and Cortona, lack of emotion recognition capability, which make people feel them as ``machine". This shows the importance of exploring on the emotion recognition system.

Researches have done in this area for decades \cite{Busso_2004,Dellaert_1996,Lee_2004_2,deng_2017}. So far, most of the work has been done on the data collected in the studio environment. The data collection was well controlled, therefore the data is clean and well segmented. Besides, most of the voice-based emotion recognition research have been done targeting on speech. In real application, there are several problems that may make current developed emotion recognition system failed. First, the non-speech voice fillers, such as laugh, whimper, cry, sigh, sob and etc., has no lexicon information but contains emotion information. Sometimes people only perform non-speech voice filler to express their emotion. Second, the voice segment length may vary in a large range. Conventional feature extraction may fail under this condition. Third, some people can control their intonation and only use the lexical information to express their emotion. Acoustic feature will not work in this case.

To address these problems, we propose an ensemble framework that combines different aspects of features from audio to develop an emotion recognition system applicable in real world. The framework is evaluated on the \emph{multimodal emotion challenge} (MEC) 2017 corpus. In this study, we focus on categorical emotion recognition, which is the task defined in MEC 2017. The corpus was collected by capturing clips from films and TV programs. These clips may contain background noise and only have non-speech voice filler, which is very close to real world scenarios. The rest parts of the paper are organized as following: Section \ref{sec:related} reviews related work about emotion recognition and previous work on audio-based approach; Section \ref{sec:corpus} describes the MEC 2017 corpus; Section \ref{sec:proposed} shows our proposed approach including the sub-system and the ensemble framework; Section \ref{sec:experiments} shows the experiments results and discuss about results analysis; Section \ref{sec:conclusion} concludes our work and discusses the future work.

\imagevspace
\section{Related Work}
\label{sec:related}
\newvspace
Voice-based emotion recognition has been done for decades. \cite{Dellaert_1996} extracted prosodic features from speech and applied majority voting of subspace specialists. It was a pilot study exploring static classifier and features for speech-based emotion recognition. \cite{Lee_2004_2} built phoneme-based dependent \emph{hidden Markov model} (HMM) classifier for emotion recognition. This work indicated the speech contents was related to emotion. Both of \cite{Lee_2004_2,Schuller_2003} showed the advantage of HMM over static model. \cite{Schuller_2010} discussed the feature set for emotion recognition task. The feature sets proposed by these works showed reliable performance. \cite{jin_2015} also used lexical information besides acoustic and showed it was helpful for acoustic event identification. However, most of the work only focused on speech part rather than non-speech voice fillers. 

Deep learning techniques were emerging as new classifier in speech related machine learning area\cite{hinton_2006}. The deep learning techniques, such as \emph{deep neural network} (DNN), \emph{convolutional neural network} (CNN) and \emph{recurrent neural network} (RNN) is able to model the feature in a high dimensional manifold space. DNN as static classifier and RNN as dynamic classifier showed their advantage in emotion recognition task compared with conventional approaches \cite{mirsamadi_2017}. Especially, \cite{mirsamadi_2017} used attention based weighted pooling to extract acoustic representation. It showed advantage over the conventional hand crafted features. Multi-task learning recently raised as a technique helping train better model for main task \cite{xia_2017,parthasarathy_2017_1}. However these work used valence and arousal as auxiliary tasks which may be difficult to get.

In this study, we use deep learning techniques with multi-task learning to build better classifier for categorical emotion recognition. The work includes the following novelties. 1. We combine different features including classical hand-craft feature and high level representation learned from deep learning techniques. The framework considered both speech and non-speech audio. 2. multi-task learning techniques were applied to DNN and weighted pooling RNN with the auxiliary tasks of speaker and gender classification, whose labels can be easily acquired. 3. Lexical information from speech was also incorporated into the system. 4. the framework was targeting at the corpus which is close to real world scenarios.

\imagevspace
\section{Corpus Description}
\label{sec:corpus}
\begin{figure}[tb]
	\centering
	{
		\includegraphics[width=\columnwidth,height=4cm]{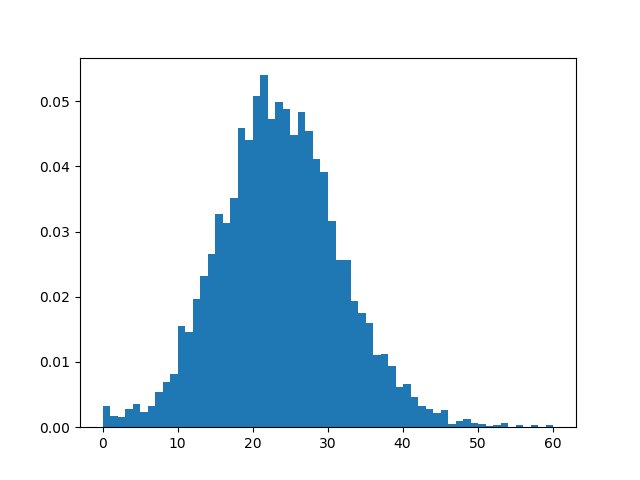}
	}\hspace{0.01mm}
	\imagevspace
	\caption{The SNR distribution of MEC 2017 corpus. The y-axis is the proportion. The x-axis is the SNR value (dB).}
	\label{fig:snr}
\end{figure}

\newvspace
This study uses the MEC 2017 corpus \cite{li_2016}. The corpus includes the clips from Chinese films, TV plays and talk shows. The clip has both of audio and video. The video is under resolution of 1028 $\times$ 680 with 24 \emph{frames per second} (FPS). The audio is under 44.1 kHz sample rate with mono channel. In this study, we downsample the audios to 16 kHz. The total duration is 5.6 hours. Each clip has one label among eight classes, angry, worried, anxious, neutral, disgust, surprise, sad and happy. The duration of the clips vary from 0.24 secs to 46.71 secs. The average duration is around 4.1 secs. There are 2105 speakers in this corpus. The gender is almost balanced. The male to female ratio is 0.46 to 0.54. The \emph{signal-to-noise ratio} (SNR) distribution is shown in Figure \ref{fig:snr}. The clips is not captured in the controlled studio environment, so there might be background noise in the clips (several clips have lower than 10 db SNR).

Since this is a challenge task, the training and testing set has been determined. The statistics for each category is listed in the Table \ref{tab:stats}. It can be seen that the data is imbalance but the distribution in training and testing sets are very similar.

\begin{table}
	\fontsize{6}{9}\selectfont
	\centering
	\caption{The distribution of MEC 2017 corpus on train set.
	}
	\begin{tabular*}{0.65\columnwidth}{@{\extracolsep{\fill}}c||c|c|c|c}
		
		\hline
		\hline
		\multirow{2}{*}{Category}&\multicolumn{2}{|c}{Training}&\multicolumn{2}{|c}{Testing} \\
		\cline{2-5}
		 & Count & Proportion & Count & Proportion \\
		\hline
		Angry & 884 & 0.180 & 128 & 0.181 \\
		Worried & 567 & 0.115 & 81 & 0.115 \\
		Anxious & 457 & 0.093 & 66 & 0.093 \\
		Neutral & 1400 & 0.285 & 200 & 0.283 \\
		Disgust & 144 & 0.029 & 21 & 0.030 \\
		Surprise & 175 & 0.036 & 25 & 0.035 \\
		Sad & 462 & 0.094 & 67 & 0.095 \\
		Happy & 828 & 0.168 & 119 & 0.168 \\
		\hline
		Total & 4917 & 1.000 & 707 & 1.000 \\
		\hline
	\end{tabular*}
	\label{tab:stats}
\end{table}

\imagevspace
\section{Proposed Approach}
\label{sec:proposed}
\begin{figure}[tb]
	\centering
	{
		\includegraphics[width=\columnwidth]{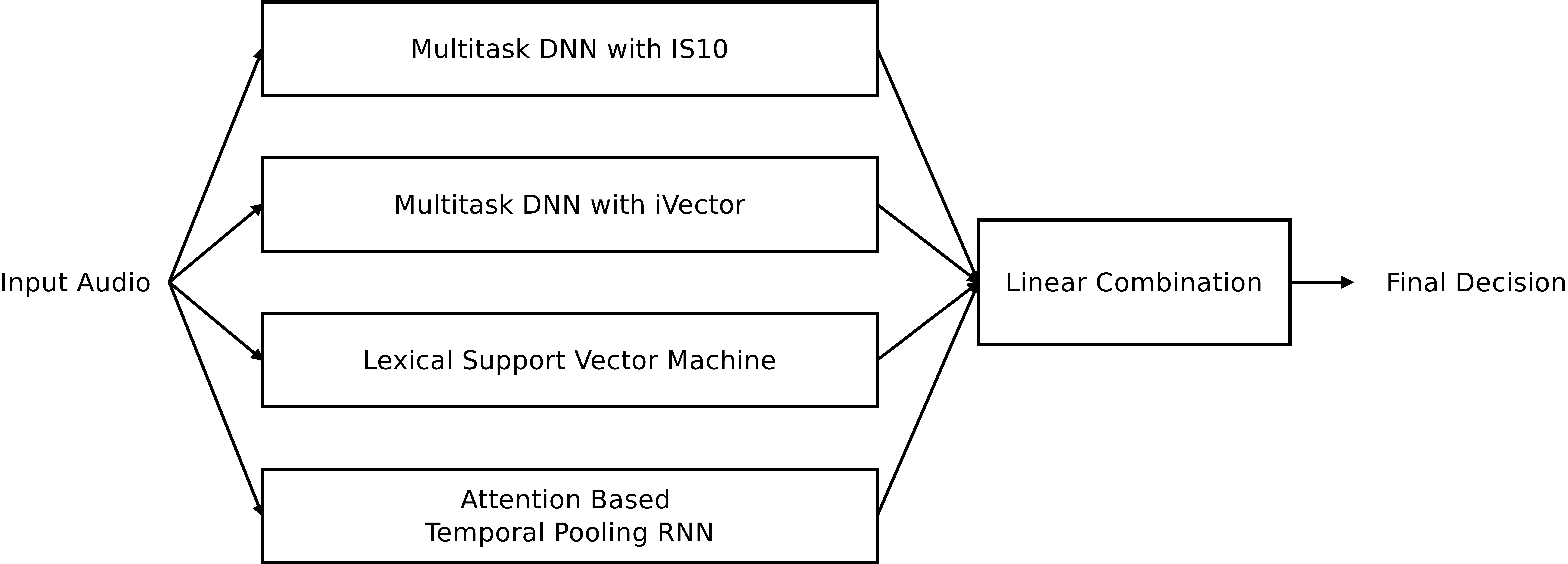}
	}\hspace{0.01mm}
	\imagevspace
	\caption{The proposed framework diagram. It has 4 sub-systems. The final decision is made by linearly combining sub-systems' decisions.}
	\label{fig:framework}
\end{figure}

\newvspace
We propose 4 sub-systems in this ensemble framework. The decisions are fused with linear combination. We use the open source toolkit Focal \cite{nikobrummer_2017} to fuse the decisions from the 4 sub-systems. It determines the linear combination weight by minimizing \emph{cost of the log-likelihood ratio} ($C_{llr}$). The framework diagram is shown in Figure \ref{fig:framework}

\newvspace
\subsection{Multi-task DNN}
\begin{figure}[tb]
	\centering
	{
		\includegraphics[width=\columnwidth]{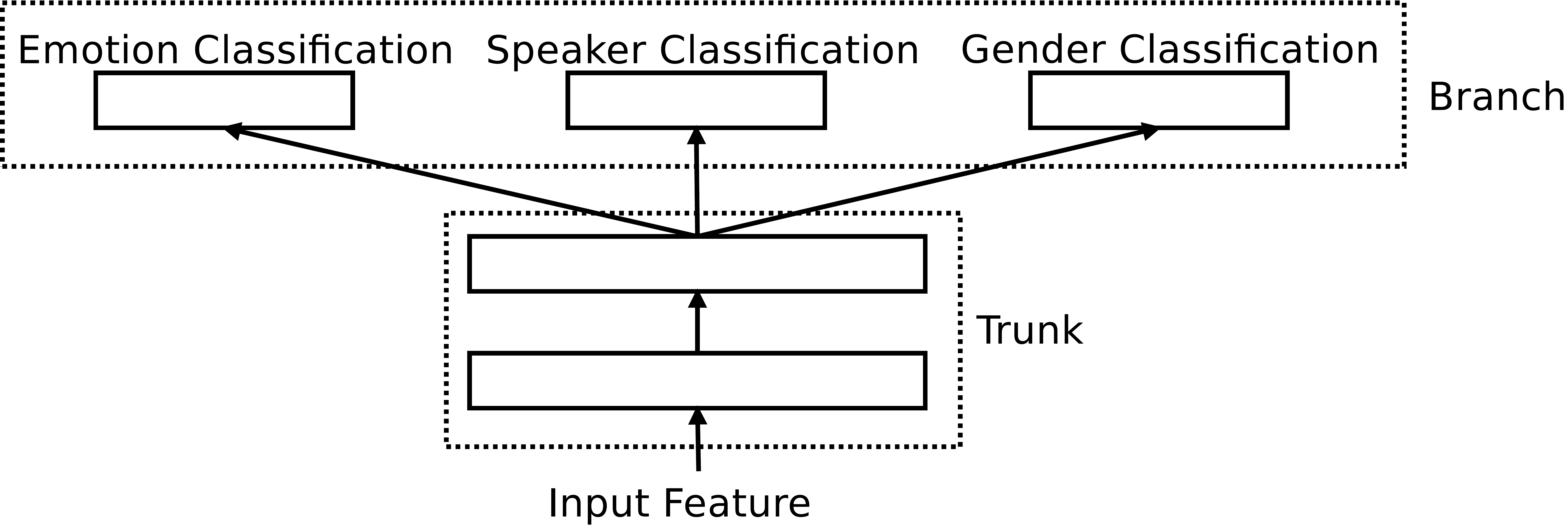}
	}\hspace{0.01mm}
	\imagevspace
	\caption{The multi-task DNN. The trunk part has the layers that are shared by all the tasks. On top of the trunk part, there is branch part for tasks. The main task is emotion classification. The auxiliary tasks are speaker and gender classifications.}
	\label{fig:dnn}
\end{figure}

\newvspace
We built multi-task DNN with the main task of emotion classification and the auxiliary tasks of speaker and gender classification. The system diagram is shown in Figure \ref{fig:dnn}. The assumption that emotion is related to speaker and gender inspires us to incorporate speaker and gender information into the classifier.

The feature set provided in Interspeech 2010 paralinguistic challenge (we name it as ``IS10" feature set) \cite{Schuller_2010} has been used and proved to work well in emotion recognition and speaker ID tasks. Therefore, we select this as one feature set to multi-task DNN. We use openSMILE \cite{Eyben_2010_3} to extract the IS10 feature set. iVector \cite{dehak_2010} has been proved to work well in speaker ID task. It was also used in emotion classification \cite{xia_2016}. Compared with IS10 feature set, it is a high level feature, which contains speaker and channel information. It is also selected as input feature to another multi-task DNN. An iVector extractor is trained based on 2000 hours of cellphone data \cite{liu_2017}. For each utterance, a 200-dimension iVector is extracted. iVector has a disadvantage that it may not be reliable with short utterance, while IS10 is not affected by utterance duration. We expect the iVector and IS10 feature set can complement each other. 

In this study, we define the architectures of the multi-task DNNs according to the inputs, since they have different dimension number. For the IS10 input, the multi-task DNN has two hidden layers in the trunk part with 4096 neurons in each layer. In the branch part, there are one hidden layer for each task with 2048 neurons. On top of that, there is a softmax layer for task classification. For the iVector input, the architecture is same. But in the trunk part the neuron number is 1024 per layer and in the branch part the neuron number is 1024 per layer. All the neuron type is \emph{rectified linear unit} (RELU). The dropout rate is 0.5.

\newvspace
\subsection{Lexical Support Vector Machine}
\newvspace
We also built a sub-system which is a \emph{support vector machine} (SVM) based on lexical information. An \emph{automatic speech recognition} (ASR) system trained with 4000 hours data was used to recognize the speech contents. The ASR is a HMM framework with \emph{time-delay DNN} (TDNN) acoustic model \cite{peddinti_2015}. After recognizing text, LibShortText toolkit is adopted for text based emotion classification \cite{yu_2013} with the feature of \emph{term frequency inverse document frequency} (TF-IDF). The classification is based on support vector classification by Crammer and Singer. 
\newvspace
\subsection{Attention Based Weighted Pooling RNN}
\begin{figure}[tb]
	\centering
	{
		\includegraphics[width=\columnwidth]{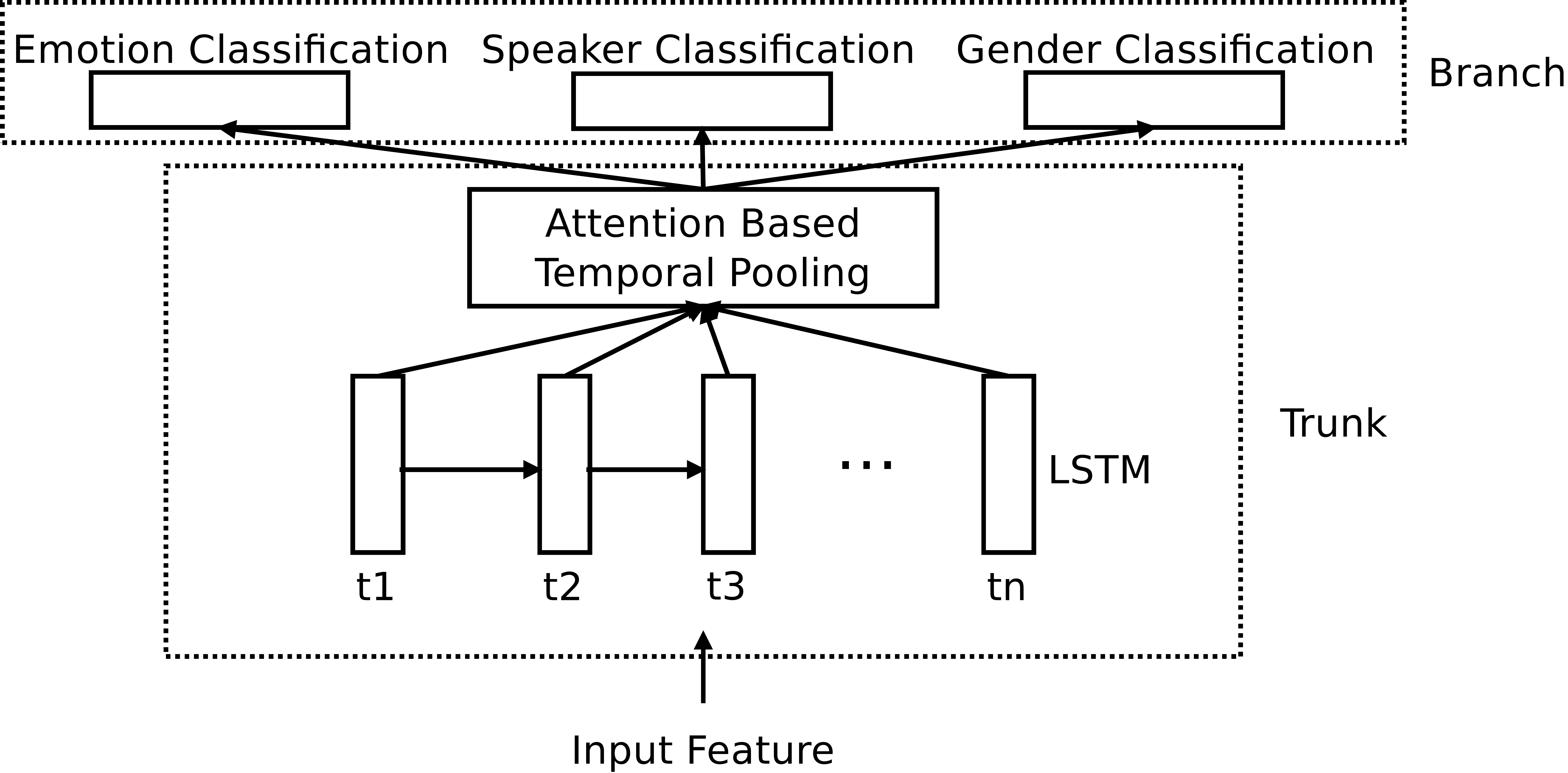}
	}\hspace{0.01mm}
	\imagevspace
	\caption{The attention based weighted pooling RNN. The LSTM layer is unrolled along the time axis (time t1 to tn). The trunk part has the layers that are shared by all the tasks. On top of the trunk part, there is branch part for tasks. The main task is emotion classification. The auxiliary tasks are speaker and gender classifications.}
	\label{fig:rnn}
\end{figure}

\newvspace
There are utterances in the corpus only containing non-speech voice filler such as laugher, whimper, sob and etc. In additional, the utterance may contain long silence or pause. Using IS10 feature set may not accurately represent acoustic characteristics. In this study, we build a sub-system which is a RNN with attention based weighted pooling to address these issues. RNN was used to model acoustic event \cite{Parascandolo_2016,Cakir_2017}, which can be voice filler. Attention based weighted pooling has been proved to work better than basic statistics, like averaging, summation and so on, because it is able to capture the informative section rather than silence or pause part \cite{mirsamadi_2017}. The system diagram is shown in Figure \ref{fig:rnn}. 

The input feature is a 36D sequential acoustic feature. The acoustic feature includes 13D MFCCs, \emph{zero crossing rate} (ZCR), energy, entropy of energy, spectral centroid, spectral spread, spectral entropy, spectral flux, spectral rolloff, 12D chroma vector, chroma deviation, harmonic ratio and pitch. It is extracted from a 25 ms window. The window shifting step size is 10 ms.

The weighted pooling is obtained by the Equation \ref{eq:temp_pool}, where $h_T$ is the hidden value output from the \emph{long short-term memory} (LSTM) layer at time $T$, and $A_T$ is the weight. $T$ is from time step $t1$ to $tn$ (sequence length is n). $A_T$ is a scalar computed by Equation \ref{eq:temp_weight}, where $W$ is the weight to be learned. Equation \ref{eq:temp_weight} is a softmax-like equation which is similar to attention model \cite{bahdanau_2016,mirsamadi_2017}. By learning $W$, it is expected that segments of interest are assigned high weight, while segments of silence or pause are assigned low weight. This model is targeting at not only speech but also other acoustic event, like voice fillers. In this model, we also use multi-task learning, which is expected to learn better representation from weighted pooling compared with single task.

In this study, we used a new type of LSTM, which is named as \emph{advanced LSTM} (A-LSTM). It is verified that it has better capability of modeling timing dependency. The A-LSTM architecture is proposed in \cite{tao_2018}.

This network has two hidden layers in the trunk part. The first one is fully connected layer with 256 RELU neurons. The second one is a \emph{bidirectional LSTM} (BLSTM) layer with 128 neurons. Weighted pooling is performed on top of the LSTM layer. The representation from weighted pooling is then sent to the branch part. In the branch part, each task has one fully connected layer with 256 RELU neurons. On top of that, there is a softmax layer performing classification. The dropout rate is 0.5.

\begin{equation} 
\label{eq:temp_pool}
Weighted Pooling = \sum_{T=t1}^{tn} A_T \times h_T
\end{equation}

\begin{equation} 
\label{eq:temp_weight}
A_T = \dfrac{\exp(W \cdot h_T)}{\sum_{T=t1}^{tn}\exp(W \cdot h_T)}
\end{equation}

\imagevspace
\section{Experiments and Results}
\label{sec:experiments}
\newvspace
The proposed approach is evaluated on MEC 2017 corpus. We build two baseline systems for comparison. One is random forest and the other one is a DNN with single task of emotion classification. Both of the baseline systems take IS10 feature set. The details of experiment settings and results are described in the following sections.

\newvspace
\subsection{Experiments Setting}
\label{sec:exp_set}
\newvspace
The training data is 4.4 hours and testing data is 0.7 hour. The batch size for all the neural network training in this study is 32. The multi-task DNN was trained with SGD optimizer, while the RNN was trained with Adam optimizer. The task weights for emotion, speaker, gender classification were 1, 0.3, 0.6 respectively. The baseline DNN with single task has three hidden layers. There are 4096 RELU neurons in the first two, and 2048 RELU neurons in the last one. This is the same architecture as multi-task DNN except that it does not have the auxiliary tasks part. The baseline random forest has 100 trees with 10 depth. This is the setting provided by the challenge organizer \cite{li_2016}. The IS10 feature was z-normalized with the mean and variance of the training data. The sequential feature was z-normalized within utterance. During training, 10\% of training data was set as validation data. The linear combination weight used in fusion was also trained from the validation set.

In the evaluation, we use the metrics of \emph{macro average F-score} (MAF) (also called unweighted F-score) and accuracy. The MAF is computed by averaging the F-score for each class detection. The accuracy is computed by dividing correctly detected sample number divided with the total sample number. Since the data is imbalance, accuracy may not be accurate to represent the system performance and MAF will be adopted as the main metric.

\newvspace
\subsection{Evaluation Results}
\newvspace
The evaluation results are shown in Table \ref{tab:result}. The performance of baseline systems, sub-systems and proposed framework is listed.

\begin{table}
	\fontsize{6}{9}\selectfont
	\centering
	\newvspace
	\caption{The performance of varieties of systems in evaluation experiments. MAF is macro average F-score.
	}
	\begin{tabular*}{0.8\columnwidth}{@{\extracolsep{\fill}}c|c||c|c}
		
		\hline
		\hline
		System Category & Approach & MAF & Accuracy (\%) \\
		\hline
		Baseline\_1 & Random Forest & 22.3 & 40.0 \\
		Baseline\_2 & DNN & 26.4 & 40.2 \\
		\hline
		\multirow{4}{*}{Sub-system} & Multi-task DNN (IS10) & 32.4 & 44.1 \\
		 & Multi-task DNN (iVector) & 27.4 & 38.0 \\
		 & Lexical SVM & 17.7 & 30.1 \\
		 & Weighted Pooling RNN & 23.2 & 39.7 \\
		 \hline
		Proposed & Ensemble Fusion & \textbf{34.2} & 41.0 \\
		\hline
	\end{tabular*}
	\label{tab:result}
\end{table}

Comparing the baseline systems, it shows the DNN outperforms the random forest by 4.1\% (absolute difference). This shows DNN has better modeling capability even with 4.4 hours training data. 
The multi-task DNN taking IS10 feature has 6\% absolute improvement compared with the baseline DNN. It indicates that the speaker and gender classification tasks are helpful for emotion classification. The multi-task DNN taking iVector feature also outperforms the baseline DNN with 1\% absolute gain. This proves iVector can also work well in emotion classification task. The performance of lexical SVM and weighted pooling RNN is lower than the baseline DNN. For the lexical SVM, it relies on the text  information from ASR which may not be perfectly reliable. For weighted pooling RNN, the shortage of training data is a key issue. 4.5 hours training to train RNN may not be sufficient.
The performance from fusion achieve highest MAF. It shows the sub-systems in the ensemble framework are complementing each other. Compared with the best baseline system, which is the single task DNN, the ensemble framework offers 7.8\% absolute improvement (about 29.5\% relative improvement).

\imagevspace
\section{Conclusion and Future Work}
\label{sec:conclusion}
\newvspace
In this study, we proposed an ensemble framework for categorical emotion recognition. The proposed framework was evaluated on MEC 2017 corpus, whose data was close to real world scenarios. We found multi-task learning with auxiliary tasks of speaker and gender classification was helpful for emotion classification. Labels for these tasks are normally easily obtained. Fusion of different sub-systems achieved better performance. It indicates capturing different aspects of features from input audio can improve the modeling capability. Since the evaluation was done on acted data in this paper, the proposed framework need be evaluated on the data from real world in the future. Besides, more data is needed for training, which may lead to better performance.

\balance
\bibliographystyle{IEEEbib}
\bibliography{reference_TAO}

\end{document}